\documentclass[a4paper,11pt]{article}
\usepackage{pos}

\newcommand{\Pp}{\ensuremath{\text{p}}}

\newcommand{\Pq}{\ensuremath{\text{q}}}
\newcommand{\Pqbar}{\ensuremath{\overline{\text{q}}}}
\newcommand{\PGg}{\ensuremath{\gamma}}
\newcommand{\PGgstar}{\ensuremath{\PGg^*}}
\newcommand{\Plp}{\ensuremath{\ell^+}}
\newcommand{\Plm}{\ensuremath{\ell^-}}

\newcommand{\PZ}{\ensuremath{\text{Z}}}

\newcommand{\mZ}{\ensuremath{\text{m}_\text{Z}}}
\newcommand{\mW}{\ensuremath{\text{m}_\text{W}}}
\newcommand{\mll}{\ensuremath{\text{m}_{\ell\ell}}}
\newcommand{\yll}{\ensuremath{y_{\ell\ell}}}

\newcommand{\sstw}{\ensuremath{\sin^2\theta_W}}

\newcommand{\sstmsbar}{\ensuremath{\sin^2\theta_{W,\ell}^{\MSbar}(\mu)}}
\newcommand{\alphamsbar}{\ensuremath{\alpha_\text{EM}^{\MSbar}(\mu)}}
\newcommand{\MSbar}{\ensuremath{\overline{\text{MS}}}}
\newcommand{\TeV}{\ensuremath{\text{TeV}}}
\newcommand{\GeV}{\ensuremath{\text{GeV}}}
\newcommand{\powheg}{\textsc{POWHEG-BOX-V2}}
\newcommand{\afb}{\ensuremath{A_\text{FB}}}
\newcommand{\diff}{\ensuremath{\text{d}}}
\newcommand{\pt}{\ensuremath{p_{T}}}
\newcommand{\tcs}{\ensuremath{\theta_{CS}}}
\newcommand{\xfitter}{\text{xFitter}}
\newcommand{\pythia}{\textsc{pythia8}}
\newcommand{\rivet}{\textsc{Rivet}}
\newcommand{\invfb}{\ensuremath{\text{fb}^{-1}}}
\newcommand{\amcfast}{\text{aMCFast}}

\title{Probing the Weak Mixing Angle at high energies at the LHC}

\author*[a]{Federico Vazzoler}
\author[a]{Simone Amoroso}
\author[a]{Ekaterina Lipka}
\author[b,c]{Clara Lavinia Del Pio}
\author[c]{Mauro Chiesa}
\author[c]{Fulvio Piccinini}
\author[d,e]{Alessandro Vicini}

\affiliation[a]{Deutsches Elektronen--Synchrotron DESY,\\
                Notkestraße 85, D--22607 Hamburg, Germany}

\affiliation[b]{Dipartimento di Fisica, Università di Pavia,\\
                via A. Bassi 6, Pavia, Italy}
  
\affiliation[c]{Istituto Nazionale di Fisica Nucleare INFN, Sezione di Pavia,\\
                via A. Bassi 6, Pavia, Italy}

\affiliation[d]{Dipartimento di Fisica, Università degli Studi di Milano,\\
                via G. Celoria 16, Milano, Italy}

\affiliation[d]{Istituto Nazionale di Fisica Nucleare INFN, Sezione di Milano,\\ 
                via G. Celoria 16, Milano, Italy}

\emailAdd{federico.vazzoler@desy.de}

\abstract{

The electroweak mixing angle is a fundamental parameter of the theory of electroweak interactions.
Its value has been measured precisely at the $\PZ$--pole at colliders.
In this contribution, we propose to exploit measurements of Neutral--Current Drell--Yan production at the Large Hadron Collider at large invariant dilepton masses to determine the energy scale dependence (running) of the electroweak mixing angle in the $\MSbar$ renormalisation scheme, $\sstmsbar$.
Such a measurement can be used to confirm the Standard Model predictions for the $\MSbar$ running at $\TeV$ scales, and to set model--independent constraints on new states with electroweak quantum numbers.
To this end, we make use of a dedicated implementation of $\sstmsbar$ in the $\powheg$ Monte Carlo event generator, which we use to explore the potential of future analyses using the data of the LHC Run 3 and High--Luminosity.

}

\FullConference{%
  41st International Conference on High Energy physics - ICHEP2022\\
  6-13 July, 2022\\
  Bologna, Italy
}

\begin{document}
\maketitle

\section{Introduction}


The electroweak mixing angle $\sstw$ is one of the fundamental parameters of the Standard Model (SM).
In the electroweak (EW) theory it can be defined as
\begin{equation}
    \label{eq:ema tree level and on shell}
    \sstw\equiv\frac{g_1^2}{g_1^2+g_2^2}=1-\frac{\mW^2}{\mZ^2},
\end{equation}
where $g_1$ and $g_2$ are the $U(1)_Y$ and $SU(2)_L$ gauge couplings, $\mW$ and $\mZ$ represent the physical masses of the massive gauge bosons, and the last equality is assumed to all orders in perturbation theory in the on--shell renormalization scheme.
In other renormalization schemes, the definition of Eq.~\ref{eq:ema tree level and on shell} is modified by radiative corrections. 
In particular, in the modified minimal--subtraction ($\MSbar$) renormalisation scheme, $\sstw$ becomes a scale--dependent quantity (running).


At hadron colliders, $\sstw$ is measured by studying the angular distribution of the lepton pairs produced in neutral current Drell--Yan (NCDY) processes $\Pq\Pqbar\to\PZ/\PGgstar\to\Plp\Plm$.
The next--to--leading (NLO) order QCD prediction of the differential cross section for the NCDY process can be expressed as a function of the angular distribution of the produced leptons, their invariant mass, $\mll$, and rapidity, $\yll$, as~\cite{Mirkes:1994dp}
\begin{equation} 
    \label{eq:differential angular Drell-Yan cross section at NLOQCD}
    \frac{\diff\sigma}{\diff\mll\diff\yll\diff\cos\tcs}=\frac{\diff\sigma_\text{unpol}}{\diff\mll\diff\yll}\biggr[(1+\cos^2\tcs)+\frac{A_0}{2}(1-3\cos^2\tcs)+A_4\cos\tcs\biggr],
\end{equation}
where $\tcs$ is the angle between the incoming and outgoing fermions in the Collins--Soper reference frame \cite{Collins:1977iv}.
Because of the vector-- and axial--vector couplings of the $\PZ$ boson, Eq.~\ref{eq:differential angular Drell-Yan cross section at NLOQCD} shows an asymmetry in the number of events expected between the forward ($\cos\tcs > 0$) and backward ($\cos\tcs < 0$) directions ($\afb$).
The value of the EW mixing angle at the scale of the $\PZ$--boson mass $\mZ$ has been extracted with a precision at the sub--per cent level~\cite{ParticleDataGroup:2022pth} by studying the $\afb$ asymmetry, while its value has never been tested at high energies, i.e. above $\mZ$.

The main goal of this work is to investigate the sensitivity of the Large Hadron Collider (LHC) data to the energy scale dependence of the EW mixing angle in the $\MSbar$ scheme above $\mZ$.
To do so, an $\MSbar$ definition of the EW mixing angle $\sstmsbar$~\cite{Erler:2004in}, where $\mu$ is chosen as the invariant mass of the final state leptons $\mu=\mll$, is considered.
This quantity is related to the coupling--based definition in Eq.~\ref{eq:ema tree level and on shell}, and it features a well--defined subtraction of the singular terms which arises in dimensional regularisation~\cite{Erler:2004in}.
To extract the expected sensitivity on $\sstmsbar$, predictions for the NCDY differential cross section of Eq.~\ref{eq:differential angular Drell-Yan cross section at NLOQCD} are obtained by generating $\Pp\Pp\to\Plp\Plm$ Monte Carlo events at a centre--of--mass energy of $13~\TeV$ in several $\mll$ bins.
We show that the LHC Run 3 data could be already sensitive to the running of $\sstmsbar$ at the per cent--level and even stronger sensitivity can be expected in the High--Luminosity LHC (HL--LHC).

\section{Simulation setup}\label{sec:simulation setup}

In the simulations used in this work, the running of $\sstmsbar$ is accounted for by using a hybrid EW scheme with $\MSbar$ input parameters $(\alphamsbar,\sstmsbar)$ and the $\PZ$ boson mass, $\mZ$, renormalised in the on--shell scheme.
The renormalisation scale $\mu$ is taken as the lepton--pair invariant mass, $\mu=\mll$.
This scheme has been introduced into an upgraded version \cite{Chiesa:XXX} of the $\textsc{Z\_ew-BMNNPV}$ code~\cite{Barze:2013fru}, within the Monte Carlo event generator framework $\powheg$~\cite{Alioli:2010xd}.

For each $\mll$ mass bin, the nominal predictions are obtained by initialising the EW parameters at their expected values at the $\PZ$ boson pole corresponding to $\mu=\mZ=91.1876\,\mathrm{GeV}$, $\alphamsbar=0.007297353$, and $\sstmsbar=0.23122$, and evolving those to the specific energy scale considered.
Additional predictions, labelled as templates, are generated by shifting the expected input value of $\sstmsbar$ up and down by a factor of $0.01$.
The values of $\mll$ and $\alphamsbar$ are modified accordingly to match their $\MSbar$ values at the corresponding energy scale.
All the predictions are obtained at LO EW and NLO+PS QCD by using the NNPDF31\_nnlo\_as\_0118\_hessian PDF set~\cite{Buckley:2014ana}, while the factorisation and renormalisation scales are chosen to be equal to $\mll$.
Events are showered with $\pythia$~\cite{Sjostrand:2014zea} without including the effect of QED radiation thus only born--level leptons are considered.

One billion nominal events are generated, separately for the electron and muon decay channel of the $\PZ$ boson, and divided in 6 $\mll$ and 6 $|\yll|$ bins separately for $\cos\tcs>0$ and $\cos\tcs<0$.
The boundaries in $\mll$ are at $116$, $150$, $200$, $300$, $500$, $1500$ and $5000\,\mathrm{GeV}$, while in $|\yll|$ are at $0.0$, $0.4$, $0.8$, $1.2$, $1.6$, $2.0$ and $2.5$.
In order to reproduce a realistic measurement scenario, the following kinematical selection criteria, inspired by the recent CMS work~\cite{CMS:2018ktx}, are imposed to the leading (sub--leading) leptons: $\pt^\ell>40\,(30)\,\GeV$ and $|\eta^\ell|<2.5$. 
Detector efficiencies and resolutions are emulated by passing each event through a dedicated $\rivet$~\cite{Buckley:2010ar} routine.
The identification and reconstruction efficiencies and energy--smearing functions used for the electrons and muons are inspired by those derived by ATLAS during the LHC Run 2 data take~\cite{ATLAS:2019jvq, ATLAS:2015lne}.

\section{Analysis strategy}\label{sec:analysis strategy}

The expected sensitivity to the running of the EW mixing angle is obtained with the $\xfitter$ analysis tool~\cite{Alekhin:2014irh}.
The templates obtained in Sec.~\ref{sec:simulation setup}, are fitted to the nominal differential cross section $\diff\sigma/\diff|\yll|\diff\mll$, simultaneously for forward ($\cos\tcs>0$) and backward ($\cos\tcs<0$) events of the nominal sample.
The expected sensitivity on the EW mixing angle in each $\mll$ bin is obtained by minimising the function
\begin{equation}
    \label{eq:chi square}\chi^2(\beta_\mathrm{exp},\beta_\mathrm{th})=\sum_{i=1}^{N_\mathrm{data}}\frac{(\sigma_i^{\mathrm{exp}}+\sum_j\Gamma_{ij}^\mathrm{exp}\beta_{j,\mathrm{exp}}-\sigma_i^\mathrm{th}-\sum_k\Gamma_{ik}^\mathrm{th}\beta_{k,\mathrm{th}})^2}{\Delta_i^2}+\sum_j\beta_{j,\mathrm{exp}}^{2}+\sum_k\beta_{k,\mathrm{th}}^2,
\end{equation}
where $\beta_\text{th.}$ are the expected theoretical uncertainties and $\beta_\text{exp.}$ the expected experimental uncertainties.
For each $\mll$ bin, the post--fit value $\delta\sstmsbar^\text{post--fit}$ is identified as the expected sensitivity on the value of the EW mixing angle at a particular value of $\mu$.

\subsection{Measurement uncertainties}\label{sec:measurements uncertainties}

Several sources of uncertainties are considered in the fitting procedure.
The expected statistical uncertainties in the data are obtained from the reconstructed--level number of events in each $(\mll,|\yll|)$ bin. 
An expected luminosity of $300\,\invfb$ ($3000\,\invfb$) is used for the Run 3 (HL--LHC) scenarios.
The systematic uncertainties are inspired by an $8~\mathrm{TeV}$ ATLAS measurement of high--mass Drell--Yan cross sections~\cite{ATLAS:2016gic}, and are extrapolated to match the $(\mll,|\yll|)$ binning used in this work by assuming a reduction factor of $0.5$ ($0.25$) for the Run 3 (HL--LHC) scenarios.
For what concerns the estimated uncertainty in the luminosity measurement, a $1\%$ uncertainty is considered for both the Run 3 and HL--LHC scenarios.
Theoretical uncertainties originating from the uncertainties in the PDFs, are modelled using $\amcfast$ grids~\cite{Bertone:2014zva, Alwall:2014hca}, convoluted with the NNPDF31\_nnlo\_as\_0118\_hessian PDF set, and included as nuisance parameters in Eq.~\ref{eq:chi square}.

\section{Results and summary}\label{sec:results}

The expected sensitivity on the running of the EW mixing angle $\delta\sstmsbar^\text{post--fit}$, defined in Sec.~\ref{sec:analysis strategy}, is presented in Tab.~\ref{tab:results}.
It ranges from about $1\%$ ($0.5\%$) to $7\%$ ($3\%$) when considering the LHC Run 3 (HL--LHC) scenario.
Due to the reduced expected statistical and systematic uncertainties, $\delta\sstmsbar^\text{post--fit}$ is lower in the HL--LHC scenario with respect to the Run 3 one. 
By repeating the fit without including the PDF uncertainties, the sensitivity in the last $\mll$ bin decreases to about $6\%$ ($2\%$) indicating a non--negligible impact of the PDF uncertainty.
\begin{table}[h]
\begin{center}
\begin{tabular}{cccc|cc}
 &  & \multicolumn{2}{c|}{Run 3 scenario} & \multicolumn{2}{c}{HL--LHC scenario} \\
\multicolumn{1}{c|}{$\mll\,[\GeV]$} & \multicolumn{1}{c|}{$\sstmsbar$} & \multicolumn{1}{c|}{$\delta\sstmsbar^\text{post--fit}$} & $[\%]$ & \multicolumn{1}{c|}{$\delta\sstmsbar^\text{post--fit}$} & $[\%]$ \\ \hline
\multicolumn{1}{c|}{133} & \multicolumn{1}{c|}{0.2334} & \multicolumn{1}{c|}{0.0021} & 0.9 & \multicolumn{1}{c|}{0.0013} & 0.5 \\
\multicolumn{1}{c|}{175} & \multicolumn{1}{c|}{0.2347} & \multicolumn{1}{c|}{0.0027} & 1.2 & \multicolumn{1}{c|}{0.0017} & 0.7 \\
\multicolumn{1}{c|}{250} & \multicolumn{1}{c|}{0.2364} & \multicolumn{1}{c|}{0.0034} & 1.4 & \multicolumn{1}{c|}{0.0021} & 0.9 \\
\multicolumn{1}{c|}{400} & \multicolumn{1}{c|}{0.2386} & \multicolumn{1}{c|}{0.0043} & 1.8 & \multicolumn{1}{c|}{0.0029} & 1.2 \\
\multicolumn{1}{c|}{1000} & \multicolumn{1}{c|}{0.2340} & \multicolumn{1}{c|}{0.0055} & 2.2 & \multicolumn{1}{c|}{0.0039} & 1.6 \\
\multicolumn{1}{c|}{3250} & \multicolumn{1}{c|}{0.2486} & \multicolumn{1}{c|}{0.0168} & 6.8 & \multicolumn{1}{c|}{0.0077} & 3.1
\end{tabular}
\end{center}
\caption{The theoretically predicted value of the EW mixing angle in the $\MSbar$ renormalisation scheme $\sstmsbar$ and the expected sensitivity $\delta\sstmsbar^\text{post--fit}$ obtained in this work. 
The results are given as a function of the invariant mass of the final state leptons $\mll$, separately for the Run 3 and HL--LHC scenarios.}
\label{tab:results}
\end{table}

\begin{figure}[h]
    \centering
    \includegraphics[scale=0.3]{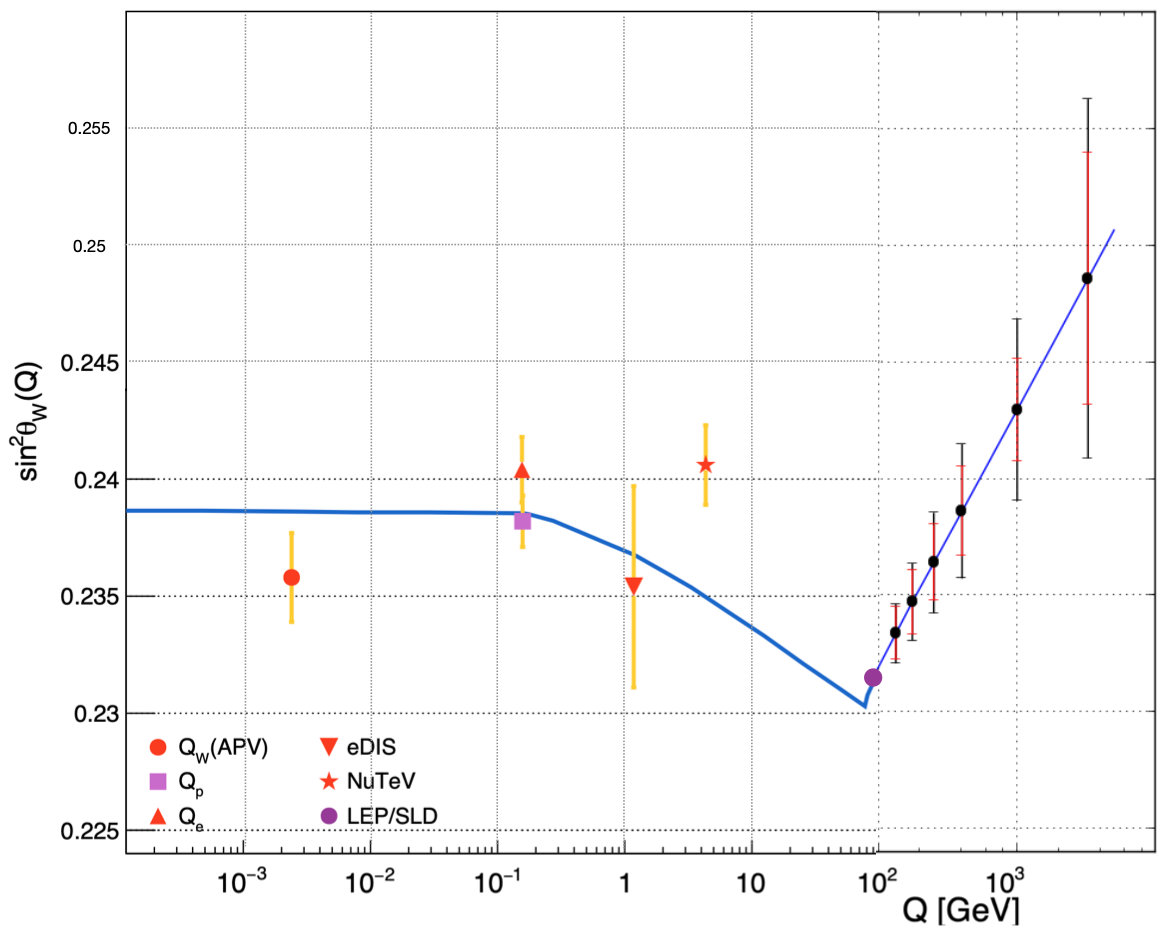}
    \caption{The running of the EW mixing angle compared to the available measurements at low energy (red points), obtained in atomic parity violation, neutrino--and polarised electron scattering experiments and to the measured value at $\mZ$ (violet point) obtained at collider experiments.
    The expectations for the HL--LHC reaches obtained in this work are shown with black points.
    The total expected uncertainties are represented as black error bars while the superimposed red error bars are obtained by repeating the fit without including the PDFs uncertainties (figure updated from~\cite{Erler:2019hds}).}
    \label{fig:results}
\end{figure}

The expected sensitivities at high energies, obtained in the HL--LHC scenario, are compared to the available measurements at and below the $\PZ$ boson pole in Fig~\ref{fig:results}.
All the measurements included in Fig~\ref{fig:results} have been translated to the $\MSbar$ renormalisation scheme~\cite{Erler:2019hds}, while the curve represents the predicted scale dependence of the EW mixing angle in the $\MSbar$ scheme obtained by including higher orders~\cite{Erler:2004in}.
This work shows that the HL--LHC will be able to probe the running of $\sstw$ at the $\TeV$ scales and to set model--independent constraints on new states with electroweak quantum numbers.

\end{document}